\documentclass[
    pre,
    aps,
    amsmath,
    amssymb,
    reprint,
    superscriptaddress,
    nofootinbib
]{revtex4-2}

\usepackage{graphicx}
\usepackage{dcolumn}
\usepackage{bm}
\usepackage[
    bookmarks=false,
    colorlinks=true,
    linkcolor=blue,
    citecolor=blue,
    urlcolor=blue,
]{hyperref}
\usepackage{orcidlink}
\usepackage{titlesec}

\DeclareMathOperator{\ord}{ord}
\DeclareMathOperator{\csch}{csch}

\DeclareMathOperator{\argmin}{argmin}

\definecolor{aesthetic-background}{RGB}{30, 20, 40}
\definecolor{aesthetic-blue}{RGB}{0, 174, 255}
\definecolor{aesthetic-cyan}{RGB}{90, 200, 230}
\definecolor{aesthetic-green}{RGB}{55, 196, 55}
\definecolor{aesthetic-magenta}{RGB}{249, 42, 173}
\definecolor{aesthetic-yellow}{RGB}{253, 163, 42}
\definecolor{ppt-blue}{RGB}{2, 83, 118}

\makeatletter
\newcommand*\@secondofsix[6]{#2}
\newcommand{\addtotitleformat}{%
  \@ifstar{\addtotitleformat@star}{\addtotitleformat@nostar}}
\newcommand\addtotitleformat@nostar[2]{%
  \PackageError{titlesec}{non starred form of \string\addtotitleformat\space not supported}{}}
\newcommand\addtotitleformat@star[2]{%
  \expandafter\expandafter\expandafter\expandafter
  \expandafter\expandafter\expandafter\def
  \expandafter\expandafter\expandafter\expandafter
  \expandafter\expandafter\expandafter\@currentsection@font
  \expandafter\expandafter\expandafter\expandafter
  \expandafter\expandafter\expandafter{%
    \expandafter\expandafter\expandafter\@secondofsix
       \csname ttlf@\expandafter\@gobble\string#1\endcsname}%
  \titleformat*{#1}{\@currentsection@font#2}%
}
\makeatother

\addtotitleformat*{\section}{\MakeUppercase}

\makeatletter
\renewcommand*\date[1][\Dated@name]{
    \def\@date{
        #1\today
    }
}
\makeatother

\begin{document}


\title{
%
    Extensible links in a broad class of single polymer chain models
}
\author{
    Michael R. Buche%
    \:\orcidlink{0000-0003-1892-0502}\,
}
\email{mrbuche@sandia.gov}
\affiliation{
    Materials and Failure Modeling, Sandia National Laboratories, Albuquerque, New Mexico 87185, USA
}
\author{
    Matthew J. Grasinger%
    \:\orcidlink{0000-0001-7188-0736}\,
}
\affiliation{
    Foundational Technologies Directorate, Air Force Research Laboratory, Wright-Patterson AFB, OH 45433, USA
}
\author{
    Jason P. Mulderrig%
    \:\orcidlink{0000-0002-0469-8122}\,
}
\affiliation{
    Foundational Technologies Directorate, Air Force Research Laboratory, Wright-Patterson AFB, OH 45433, USA
}
\affiliation{
    National Research Council (NRC) Research Associateship Programs, The National Academies of Sciences, Engineering, and Medicine
}
\date{}

\begin{abstract}
The physics of polymer chains is often probed using molecular stretching experiments and various idealized single-chain models.
The majority of these models consist of a discrete sequence of links, which may be treated as rigid or extensible.
Although such models are well established and many specific extensible variants have been proposed, no generally applicable theory has been presented.
Moreover, most existing treatments are heuristic rather than systematically and rigorously derived.
This critical gap is closed here through the development of a generally applicable asymptotic theory for including link extensibility in a broad class of discrete models for single-chain thermodynamics.
The theory is verified analytically using the freely jointed chain and validated numerically using the freely rotating chain.
The resulting approximation is first-order accurate in inverse link stiffness, with quadratically decreasing error, and recovers extensible behavior across all link stiffnesses from a single rigid-link reference calculation.
\smallskip\smallskip\smallskip

\noindent DOI: \href{https://doi.org/10.1103/h5wp-gmp6}{10.1103/h5wp-gmp6}
\end{abstract}

\maketitle


Building a fundamental understanding of the physics of single polymer chains is an essential first step toward understanding polymeric materials more generally.
One important aspect of single-chain polymer physics is the behavior at thermal equilibrium, which is described by statistical thermodynamics \cite{mcquarrie}.
At thermal equilibrium, a chain resists extension in order to maximize entropy \cite{kuhn1942beziehungen}, where this behavior is described by a partition function for that particular statistical ensemble \cite{manca2012elasticity,buche2020statistical}.
Idealized single-chain models are typically applied in evaluating these partition functions to make them computationally tractable or analytically solvable \cite{wang1952statistical}, the most common of which being the freely jointed chain model \cite{treloar1949physics}.
Since most of these models are simply a discrete sequence of rigid links as originally formulated \cite{rubinstein2003polymer}, they are unable to predict chemical bond disruption.
This is a vital gap, since most models for polymer chains \cite{mao2017rupture,buche2022freely,mulderrig2023statistical} or networks \cite{buche2021chain,mulderrig2021affine,lamont2021rate} which have chain scission are typically predicated on bond stretch.
Moreover, extensibility must be included when modeling single-chain stretching experiments \cite{buche2023modeling} with larger applied forces \cite{zhang2025enhancing}.
Considerable effort has been devoted to these problems \cite{balabaev2009extension,radiom2017influence,mao2017rupture,buche2026link,fiasconaro2019analytical,lavoie2019modeling,buche2022freely,mulderrig2023statistical,fiasconaro2023elastic,zhu2025stretching}, but no generally applicable rigorous theory has yet emerged.

The majority of work thus far has sought to replace the rigid links with their extensible counterparts through a potential energy function \cite{buche2022freely}.
For a freely jointed chain, the resulting extensible model can be either analytically approximated \cite{fiasconaro2019analytical,buche2022freely} or exactly analytically solved \cite{balabaev2009extension}.
However, in the case of the freely rotating chain only phenomenological \cite{fiasconaro2023elastic} or effective \cite{zhu2025stretching} contributions from link extensibility have been developed.
Critically, these approximations obscure underlying physics and lack any error estimation, both of which would accompany a truly rigorous derivation.
Moreover, no reliable theory exists for replacing link rigidity with extensibility in general.
Instead, existing approaches largely rely on heuristically importing ideas from mechanics \cite{mao2017rupture,lavoie2019modeling}. To close this gap, an asymptotically correct theory is developed and verified here for incorporating link extensibility into a broad class of discrete single-chain models.

Many single-chain polymer models consist of a discrete series of identical links.
Most of these are governed by two separable potential energies, where the first $U_0$ is a function of the link angles and the second $U_1$ is a function of the link lengths.
This separability is widely applicable to idealized single-chain models, including the examples considered here, but is not typically applicable to fully atomistic molecular dynamics models.
The total potential energy $\Pi$ in the isotensional ensemble would then be
\begin{equation}\label{eq:pi}
\Pi(\boldsymbol{\ell},\boldsymbol{\theta},\boldsymbol{\phi}) = U_0(\boldsymbol{\theta},\boldsymbol{\phi}) + U_1(\boldsymbol{\ell}) - \mathbf{f}\cdot\boldsymbol{\xi},
\end{equation}
where $\boldsymbol{\ell}$ represents the link lengths, $\boldsymbol{\theta}$ the polar angles, $\boldsymbol{\phi}$ the azimuthal angles, $\mathbf{f}$ the applied force vector, and $\boldsymbol{\xi}$ the chain end-to-end vector.
If there are $N_b$ links in the chain, the potential $U_1$ is merely the sum of the individual link potentials $u(\ell_i)$ for $i=1,\ldots,N_b$.
Assuming these are predominantly harmonic, the nondimensional form is
\begin{equation}\label{eq:u1}
\beta U_1(\boldsymbol{\lambda}) = \frac{\kappa}{2}\sum_{i=1}^{N_b}\left(\lambda_i - 1\right)^2,
\end{equation}
where $\boldsymbol{\lambda}=\boldsymbol{\ell}/\ell_b$ are the link stretches, $\kappa=\beta k_b\ell_b^2$ is the nondimensional link stiffness, and $\beta=1/kT$ as usual \cite{buche2022freely}.
If $\eta=\beta f\ell_b$ is the nondimensional force applied along the polar axis, the nondimensional form of Eq.~\eqref{eq:pi} is
\begin{equation}\label{eq:pi-n}
\beta\Pi(\boldsymbol{\lambda},\boldsymbol{\theta},\boldsymbol{\phi}) = \beta U_0(\boldsymbol{\theta},\boldsymbol{\phi}) + \beta U_1(\boldsymbol{\lambda}) - \eta\sum_{i=1}^{N_b}\lambda_i\cos\theta_i.
\end{equation}
The isotensional partition function $Z(\eta)$ is calculated by integrating $\exp(-\beta\Pi)$ over all configurations $(\boldsymbol{\ell},\boldsymbol{\theta},\boldsymbol{\phi})$.
This partition function can be written as \cite{buche2021fundamental}
\begin{equation}\label{eq:z}
Z(\eta) = \int Z_0(\eta,\boldsymbol{\lambda})\,e^{-\beta U_1(\boldsymbol{\lambda})}\,d\boldsymbol{\lambda},
\end{equation}
where the reference system partition function $Z_0(\eta,\boldsymbol{\lambda})$, which is the system with rigid links, is then given by
\begin{equation}\label{eq:z0}
Z_0(\eta,\boldsymbol{\lambda}) = \int e^{-\beta U_0(\boldsymbol{\theta},\boldsymbol{\phi}) + \eta\sum_i\lambda_i\cos\theta_i}\,d\boldsymbol{\Gamma}.
\end{equation}
Note that $d\boldsymbol{\lambda}=\prod_i d\lambda_i$ and $d\boldsymbol{\Gamma}=\prod_i\lambda_i^2\sin\theta_i\,
d\theta_i\,d\phi_i$, where $\lambda_i\in[0,\infty)$, $\theta_i\in[0,\pi]$, and $\phi_i\in[0,2\pi)$.
If there are any additional rigid constraints on the chain via the link angles, they are enforced within $U_0(\boldsymbol{\theta},\boldsymbol{\phi})$.

Now, the asymptotic theory is developed.
For $\kappa\gg 1$, the configurations predominantly contributing to Eq.~\eqref{eq:z} are tightly clustered around $\argmin\beta U_1$, which is $\boldsymbol{\lambda}=\mathbf{1}$.
Systematically applying Laplace's method for evaluating integrals \cite{bender2013advanced}, the partition function for the full system $Z(\eta)$ can be asymptotically approximated in terms of that for the reference system $Z_0(\eta)$ \cite{buche2021fundamental}.
The result is
\begin{equation}\label{eq:z-a}
Z(\eta) \sim \left(\frac{2\pi}{\kappa}\right)^{N_b/2}Z_0(\eta)\left[1 + \frac{h(\eta)}{\kappa} + \ord\left(\kappa^{-2}\right)\right],
\end{equation}
where $Z_0(\eta)\equiv Z_0(\eta,\mathbf{1})$, and $h(\eta)$ is given by \cite{buche2021fundamental}
\begin{equation}\label{eq:h}
h(\eta) = \frac{1}{2Z_0}\sum_{i=1}^{N_b}\left.\frac{\partial^2Z_0}{\partial\lambda_i^2}\right|_{\boldsymbol{\lambda}=\mathbf{1}}.
\end{equation}
The nondimensional longitudinal chain extension $\gamma(\eta)$ is the average over the chain of the ensemble average of the component of each link vector along the polar axis,
\begin{equation}\label{eq:gamma}
\gamma(\eta) = \frac{1}{N_b}\sum_{i=1}^{N_b}\left\langle\lambda_i\cos\theta_i\right\rangle = \frac{1}{N_b}\frac{\partial\ln Z}{\partial\eta},
\end{equation}
and analogously, for the reference system at $\boldsymbol{\lambda}=\mathbf{1}$,
\begin{equation}\label{eq:gamma_0}
\gamma_0(\eta) = \frac{1}{N_b}\sum_{i=1}^{N_b}\left\langle\cos\theta_i\right\rangle_0 = \frac{1}{N_b}\frac{\partial\ln Z_0}{\partial\eta}.
\end{equation}
The derivative in Eq.~\eqref{eq:h} is analytically evaluated as
\begin{equation}
\frac{1}{Z_0}\left.\frac{\partial^2Z_0}{\partial\lambda_i^2}\right|_{\boldsymbol{\lambda}=\mathbf{1}} = 2 + 4\eta\left\langle\cos\theta_i\right\rangle_0 + \eta^2\left\langle\cos^2\theta_i\right\rangle_0,
\end{equation}
and with Eq.~\eqref{eq:gamma_0} allows $h(\eta)$ in Eq.~\eqref{eq:z-a} to be written as
\begin{equation}
h(\eta) = N_b + 2N_b\eta\gamma_0(\eta) + \frac{\eta^2}{2}\sum_{i=1}^{N_b}\left\langle\cos^2\theta_i\right\rangle_0.
\end{equation}
Finally, the nondimensional longitudinal extension in Eq.~\eqref{eq:gamma} is asymptotically approximated via Eq.~\eqref{eq:z-a} as
\begin{equation}\label{eq:gamma-a}
\gamma(\eta) \sim \gamma_0(\eta) + \frac{1}{\kappa}\Big[2\gamma_0(\eta) + 2\eta\gamma_0'(\eta) + g(\eta)\Big],
\end{equation}
valid for $\kappa\gg 1$.
The function $g(\eta)$ is given by
\begin{equation}
g(\eta) = \frac{\eta}{N_b}\sum_{i=1}^{N_b}\big[a_i(\eta) + b_i(\eta)\big],
\end{equation}
where the first term does not involve cross terms,
\begin{equation}\label{eq:ai}
a_i(\eta) = \left(1 - \frac{1}{2}\,N_b\eta\gamma_0\right)\left\langle\cos^2\theta_i\right\rangle_0,
\end{equation}
but the second term involves cross terms for all links,
\begin{equation}\label{eq:bi}
b_i(\eta) = \frac{\eta}{2}\sum_{j=1}^{N_b}\left\langle\cos^2\theta_i\cos\theta_j\right\rangle_0.
\end{equation}
Also, the derivative of $\gamma_0(\eta)$ with respect to $\eta$ is
\begin{equation}\label{eq:dg0}
\gamma_0'(\eta) = \frac{1}{N_b}\sum_{i,j}\left\langle\cos\theta_i\cos\theta_j\right\rangle_0 - N_b\gamma_0(\eta)^2.
\end{equation}
The relation in Eq.~\eqref{eq:gamma-a} is asymptotically valid for adding link extensibility to all single-chain models with separable potential energy functions vis-\'a-vis Eqs.~\eqref{eq:u1} and \eqref{eq:pi-n}.
This broad class of models includes the freely jointed chain model \cite{fiasconaro2019analytical,buche2022freely}, the freely rotating chain model \cite{livadaru2003stretching,zhu2025stretching}, the hindered rotation model \cite{flory1969statistical,rubinstein2003polymer}, the rotational isometric state model \cite{rubinstein2003polymer,helfer2007rotational}, and discrete representations of the worm-like chain model \cite{becker2010radial,manca2012elasticity,fiasconaro2023elastic}.
For the sake of brevity, two single-chain models are selected to demonstrate the asymptotic theory.
First, Eq.~\eqref{eq:gamma-a} is verified using the freely jointed chain model, which is exactly solvable in both the inextensible and extensible cases.
Second, Eq.~\eqref{eq:gamma-a} is validated using the freely rotating chain model, which is not analytically solvable in either the inextensible or the extensible case.

\begin{figure}[t]
\includegraphics{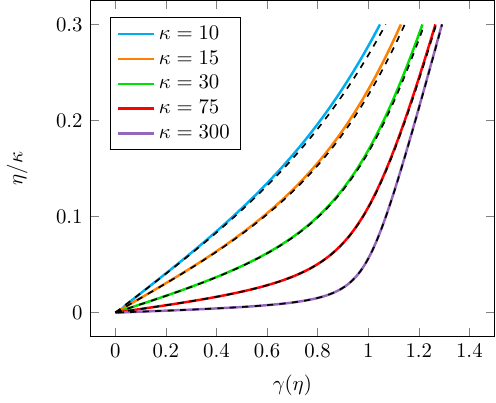}
\caption{\label{fig:efjc}%
The nondimensional longitudinal extension $\gamma$ of an extensible freely jointed chain under an applied force $\eta$, using asymptotic (dashed) and analytically-exact (solid) methods, for varying nondimensional link stiffness $\kappa$.
}
\end{figure}

For the inextensible freely jointed chain model \cite{buche2022freely}, the nondimensional longitudinal extension is given by the Langevin function \cite{kuhn1942beziehungen}, $\gamma_0(\eta)=\mathcal{L}(\eta)=\coth\eta-\eta^{-1}$.
The derivative of the Langevin function is then given by $\gamma_0'(\eta)=\eta^{-2} - \csch^2\!\eta$.
Each link in a freely jointed chain under force is independent and identical \cite{buche2026fluctuations,buche2026link}, so
\begin{equation}\label{eq:fjc:cos2}
\left\langle\cos^2\theta\right\rangle_0 = 1 - \frac{2}{\eta}\coth\eta + \frac{2}{\eta^2}.
\end{equation}
The independence of links also causes
\begin{equation}
\left\langle\cos^2\theta_i\cos\theta_j\right\rangle_0 = \begin{cases}
\langle\cos^3\theta\rangle_0,& i=j,\\
\langle\cos^2\theta\rangle_0\langle\cos\theta\rangle_0,& i\neq j,\end{cases}
\end{equation}
where $\langle\cos\theta\rangle_0=\gamma_0(\eta)$ here, and one can compute
\begin{equation}\label{eq:fjc:cos3}
\left\langle\cos^3\theta\right\rangle_0 = \coth\eta - \frac{3}{\eta} + \frac{6}{\eta^2}\coth\eta - \frac{6}{\eta^3}.
\end{equation}
Substituting Eqs.~\eqref{eq:fjc:cos2}--\eqref{eq:fjc:cos3} into Eqs.~\eqref{eq:ai}--\eqref{eq:dg0}, and then substituting the simplified relations into Eq.~\eqref{eq:gamma-a}, yields
\begin{equation}\label{eq:gamma-efjc-a}
\gamma(\eta) \sim \mathcal{L}(\eta) + \frac{\eta}{\kappa}\Big[2 - \mathcal{L}(\eta)\coth\eta\Big],
\end{equation}
where the nondimensional link stiffness is large, $\kappa\gg 1$.
Eq.~\eqref{eq:gamma-efjc-a} is exactly the expected result here, which is the series expansion of the exact solution in powers of $\kappa^{-1}$, neglecting terms that are $\ord(\kappa^{-2})$ and beyond \cite{buche2021chain,buche2022freely}.
Fig.~\ref{fig:efjc} compares the asymptotic approximation of the longitudinal extension $\gamma(\eta)$ in Eq.~\eqref{eq:gamma-efjc-a} to the exact analytic result \cite{balabaev2009extension,buche2022freely} as a function of the nondimensional force $\eta$ across a range of nondimensional link stiffnesses $\kappa$ \cite{conspire}.
For reference, $\kappa\sim 10^2$ to $10^3$ is typical for these models \cite{buche2021chain,mulderrig2023statistical,zhu2025stretching}.
The asymptotic approximation is moderately accurate for smaller values of $\kappa$, quickly becoming more accurate as $\kappa$ increases.
For larger $\kappa$, the exact and approximate results become indistinguishable, highlighting the rapid convergence of the asymptotic theory.

In the case of the freely rotating chain \cite{livadaru2003stretching}, neither the inextensible reference system nor the actually extensible system is analytically solvable.
Therefore, Monte Carlo calculations are utilized to validate the theory instead.
Note that these calculations are much more expedient for the reference system than the full system, since the reference system is constant in energy.
Also note that the asymptotic theory extends reference system calculations to approximate the full system over a range of link stiffnesses without requiring additional calculations.

To build the asymptotic relation for $\gamma(\eta)$ in Eq.~\eqref{eq:gamma-a}, $\langle\cos^m\theta_i\cos^n\theta_j\rangle$ is calculated in the inextensible system across $m,n=1,2$ and $i,j=1,\ldots,N_b$.
Configurations of the inextensible freely rotating chain are generated by rejection-free sampling in the isometric ensemble, and averages of any quantity \(A\) in the isotensional ensemble are then obtained by exponential reweighting:
\begin{equation}\label{eq:reweighting}
\left\langle A\right\rangle_\eta = \frac{\left\langle Ae^{N_b\boldsymbol{\eta}\cdot\boldsymbol{\gamma}}\right\rangle_\gamma}{\left\langle e^{N_b\boldsymbol{\eta}\cdot\boldsymbol{\gamma}}\right\rangle_\gamma}.
\end{equation}
To calculate the exact relation for $\gamma(\eta)$ given by Eq.~\eqref{eq:gamma}, $\langle\lambda_i\cos\theta_i\rangle$ is calculated in the extensible system for all $i=1,\ldots,N_b$ links.
The configurations of the extensible freely rotating chain are generated analogously to the inextensible case, but the link lengths are randomly sampled from the Maxwell-Boltzmann distribution associated with the link stiffness.
To encourage sampling of rare highly extended states, the stretch distribution is biased in proportion to \(e^{\eta \lambda_i}\), and the resulting auxiliary ensemble is then reweighted according to
\begin{equation}\label{eq:reweighting-bias}
\left\langle A\right\rangle_\eta = \frac{\left\langle Ae^{\eta\sum_i\lambda_i(\cos\theta_i - 1)}\right\rangle_\gamma}{\left\langle e^{\eta\sum_i\lambda_i(\cos\theta_i - 1)}\right\rangle_\gamma}.
\end{equation}

\begin{figure}[t]
\includegraphics{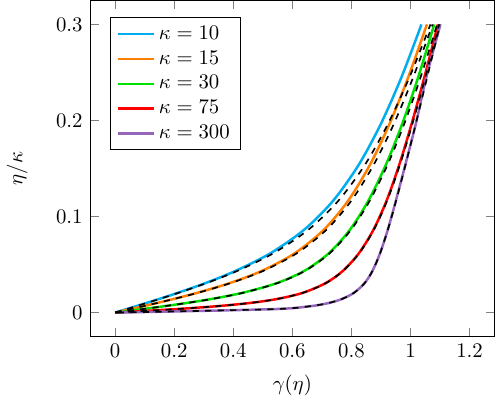}
\caption{\label{fig:efrc}%
The nondimensional longitudinal extension $\gamma$ of an extensible freely rotating chain under an applied force $\eta$, using asymptotic (dashed) and numerically-exact (solid) methods, for varying nondimensional link stiffness $\kappa$.
}
\end{figure}

Fig.~\ref{fig:efrc} compares the asymptotic approximation of the longitudinal extension $\gamma(\eta)$ in Eq.~\eqref{eq:gamma-a} to the exact result in Eq.~\eqref{eq:gamma} as a function of the nondimensional force $\eta$ across a range of nondimensional link stiffnesses $\kappa$ \cite{conspire}.
Each case averages 10 billion Monte Carlo samples for every one of the 15 points along each curve, where the angle between consecutive link vectors is fixed at $\theta_b=60^\circ$ in a freely rotating chain model consisting of $N_b=5$ links.
For the exact relation, the Monte Carlo calculations must be repeated for every $\kappa$ and $\eta$, whereas for the asymptotic relation, the same reference system calculations at a given $\eta$ can be reused for all values of $\kappa$.
The results in Fig.~\ref{fig:efrc} show that the asymptotic theory provides accurate approximations of the extensible freely rotating chain.
Similar to Fig.~\ref{fig:efjc}, the asymptotic relation is moderately accurate for smaller $\kappa$ before becoming highly accurate for larger $\kappa$.

\begin{figure}[t]
\includegraphics{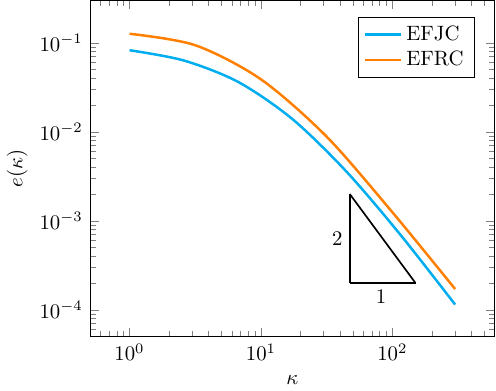}
\caption{\label{fig:error}%
The relative error $e$ in the asymptotic theory as a function of the nondimensional link stiffness $\kappa$ for both the extensible freely jointed chain (EFJC) and extensible freely rotating chain (EFRC).
}
\end{figure}

To quantify the accuracy of the asymptotic theory, the relative error $e(\kappa)$ is calculated using the $L_2$ norm \cite{buche2022freely}
\begin{equation}
e(\kappa) = \sqrt{\frac{\int_0^{\eta_0}\Delta\gamma(\eta)^2\,d\eta}{\int_0^{\eta_0}\gamma(\eta)^2\,d\eta}},
\end{equation}
where $\Delta\gamma(\eta)$ is the difference between the exact and asymptotic results.
The upper limit $\eta_0=10$ is kept fixed as $\kappa$ varies so that the error scaling reflects the accuracy over a common force interval.
Otherwise, the error scaling could reflect changes in the domain of integration and obscure the convergence rate associated with the asymptotic approximation.
At room temperature, $\eta_0=10$ corresponds to forces of roughly $40$--$400\,\mathrm{pN}$ for link lengths $\ell_b=1$--$0.1\,\mathrm{nm}$, which are in the relevant range for single-molecule force spectroscopy \cite{marko1998dna,radiom2017influence}.
Fig.~\ref{fig:error} shows the $L_2$ relative error $e(\kappa)$ in the asymptotic theory as a function of the nondimensional link stiffness $\kappa$ for both the extensible freely jointed chain (EFJC) and the extensible freely rotating chain (EFRC).
For small values of $\kappa$ such as those near $\kappa=1$, the error is considerable since the asymptotic theory is only valid for $\kappa\gg 1$.
As $\kappa$ increases, the error steadily decreases, and for large values of $\kappa$, the log-log slope of the error converges to $-2$.
This further validates the asymptotic theory through indicating that the error is indeed $\ord(\kappa^{-2})$ and the theory is therefore correct to first order for $\kappa\gg 1$.

A general asymptotic theory has been developed which permits link extensibility in a broad class of models for the statistical thermodynamics of single polymer chains under force.
The theory systematically builds accurate approximations upon the inextensible reference system.
The accuracy and broad applicability of this theory were demonstrated using two exemplary single-chain models, the freely jointed chain and the freely rotating chain.
This was done both analytically and numerically, since the former model is exactly solvable while the latter model is not.
The success of the asymptotic theory here provides promising evidence that it could apply much more generally in the future.
Specifically, it could be applied to other discrete link-based molecular models, general discrete models for statistical thermodynamics, or even models with continuous degrees of freedom.

\begin{acknowledgments}
Sandia National Laboratories is a multi-mission laboratory managed and operated by National Technology and Engineering Solutions of Sandia, LLC., a wholly owned subsidiary of Honeywell International, Inc., for the U.S. Department of Energy's National Nuclear Security Administration under Contract No. DE-NA0003525. Any subjective views or opinions expressed in the paper do not necessarily represent the views of the U.S. Department of Energy or the U.S. Government. The U.S. Government retains and the publisher, by accepting the article for publication, acknowledges that the U.S. Government retains a nonexclusive, paid-up, irrevocable, world-wide license to publish or reproduce the published form of this manuscript, or allow others to do so, for U.S. Government purposes.

MRB acknowledges the support of Sandia National Laboratories.
MG acknowledges the support of the Air Force Research Laboratory.
JPM gratefully acknowledges the support of the National Research Council (NRC) Research Associateship Program (administered by the National Academies of Sciences, Engineering, and Medicine).
\end{acknowledgments}

\bibliography{main}

\end{document}